\documentclass[aps,superscriptaddress,nofootinbib,preprintnumbers,showpacs,twocolumn,prd]{revtex4}

\usepackage{txfonts}

\usepackage{amssymb}

\usepackage{indentfirst}

\usepackage{bm}

\usepackage{epsfig}

\usepackage{epsf}

\usepackage{graphicx}

\newcommand{\be}{\begin{equation}}
\newcommand{\ee}{\end{equation}}
\newcommand{\ba}{\begin{eqnarray}}
\newcommand{\ea}{\end{eqnarray}}

\renewcommand{\d}{{\mathrm d}}

\newcommand{\kp}{\boldsymbol{k}_T}

\newcommand{\ssh}{\!\!\!/}

\renewcommand{\d}{{\mathrm d}}

\begin{document}

\title{The $\cos2\phi$ azimuthal asymmetry of unpolarized dilepton production at the $Z$-pole }

\author{Zhun Lu}
\affiliation{Department of Physics, Southeast University, Nanjing
211189, China}
\affiliation{Departamento de F\'\i sica, y
Centro Cient\'ifico-Tecnol\'ogico de Valpara\'iso, Universidad T\'ecnica Federico Santa Mar\'\i a,
Casilla 110-V, Valpara\'\i so, Chile}

\author{Ivan Schmidt}
\affiliation{Departamento de F\'\i sica, y
Centro Cient\'ifico-Tecnol\'ogico de Valpara\'iso, Universidad T\'ecnica Federico Santa Mar\'\i a,
Casilla 110-V, Valpara\'\i so, Chile}

\begin{abstract}
We calculate the Boer-Mulders effect contribution to the $\cos2\phi$
azimuthal asymmetry of unpolarized dilepton production near the
$Z$-pole. Based on the tree-level expression in the transverse
momentum dependent factorization framework, we show that the
corresponding asymmetry near the $Z$-pole is negative, which is
opposite to the asymmetry in the low $Q^2$ region, dominated by the
production via a virtual photon. We calculate the asymmetry
generated by the Boer-Mulders effect near the $Z$-pole at RHIC, with
$\sqrt{s}=500$ GeV. We find that the magnitude of the asymmetry is
several percent, and therefore it is measurable. The experimental
confirmation of this sign change of the asymmetry from the low $Q^2$
region to the $Z$-pole provides direct evidence of the chiral odd
structure of quarks inside an unpolarized nucleon.

\end{abstract}


\maketitle

\section{introduction}

The $\cos 2 \phi$ angular distribution of dilepton production in unpolarized hadron collisions
$h_1  h_2  \rightarrow \ell^+   \ell^-  +X$
belongs to the remaining challenges
which need to be understood from QCD dynamics~\cite{Barone:2010ef}.
According to the Lorentz structure of the hadronic tensor, one can write down
the differential cross-section of dilepton production as ~\cite{oakes,lt78}:
\begin{eqnarray}
{d\sigma\over \d\Omega d^4 q} &=& W_T (1+ \cos^2 \theta)   + W_L (1- \cos^2\theta) \nonumber\\
&& + W_\Delta \sin 2\theta \cos \phi  + W_{\Delta \Delta} \sin^2 \theta
\cos 2\phi \; . \label{acs}
\end{eqnarray}
Here $q$ is the virtual photon or $Z$ boson's four momentum,
and $d\Omega = d\cos\theta d\phi$ is the solid angle of the lepton $\ell$
in terms of its polar and azimuthal angles in the center-of-mass system
(c.m.s.) of the lepton pair. The coefficient functions $W_{T,L,\Delta,\Delta \Delta}$
depend on the invariant mass $Q$,  transverse momentum
$q_T$, and the rapidity $y$ of $\gamma^*/Z$. After the solid angle $d\Omega$ is integrated over,
the differential cross-section with respect to $q$ has the form
\begin{eqnarray}
{d\sigma\over d^4 q} &=& {8\pi\over 3}\left(2W_T +  W_L\right)\; . \label{qcs}
\end{eqnarray}
Therefore, the angular distribution of the lepton pair is defined as
\begin{eqnarray}
\frac{dN}{d\Omega} = \left.\frac{d \sigma}{d\Omega d^4 q}\right/\frac{d \sigma}{d^4 q} \; . \label{adis}
\end{eqnarray}
Equivalently, another often used convention for the dilepton angular distribution  is
\begin{eqnarray}
\frac{dN}{d\Omega}&=&\left.\frac{3}{4\pi}\frac{1}{\lambda+3}
\right(1+\lambda\,\cos^2\theta+\mu\,\sin2\,\theta\,\cos\phi
\nonumber\\
& & \left.+\frac{\nu}{2}\sin^2\theta\cos2\phi\right).\label{cos2phi}
\end{eqnarray}
Comparing (\ref{acs}) and (\ref{cos2phi}) yields following relations
\ba
\lambda=\frac{W_T-W_L}{W_T+W_L}\; , \;\;\;
\mu=\frac{W_\Delta}{W_T+W_L}\; , \;\;\;
\nu=\frac{2W_{\Delta\Delta}}{W_T+W_L} \; .
\label{coedif}
\ea

Of particular interest are the angular distribution given by the $\lambda$ and $\nu$ terms.
To $\alpha_s$ order of perturbative QCD, a calculation~\cite{lt79} in collinear factorization
showed that these coefficients satisfy:
\begin{eqnarray}
2\mu+\lambda-1=0\;, \label{ltre}
\end{eqnarray}
the so-called Lam-Tung relation~\cite{lt78,lt80}, which has attracted considerable attention. Fixed-order pQCD
calculations~\cite{mo94} at order $\alpha_s^2$, as well as QCD resummation
calculations~\cite{bv06,berger07} to all orders in collinear factorization, indicate that violations of
(\ref{ltre}) are very small. However, early measurements on $\pi^- N \to \gamma^*+X\to \ell^+ \ell^- +X $
processes by the NA10~\cite{na10} and E165~\cite{conway} Collaborations at $\sqrt{s}=19$ and $23\,\text{GeV}$,
show large positive values of $\nu$, near 30\%, indicating a sizable violation of the Lam-Tung relation.
The relation has also been tested in $pp$ and $pd$ Drell-Yan processes by the E866/NuSea
collaboration~\cite{e866,e866pp} at $\sqrt{s}=38.7\, \text{GeV}$, and very recently in $p\bar{p} \to   \gamma^\star/Z+ X \rightarrow l^+ l^- + X$ by the CDF Collaboration ~\cite{Aaltonen:2011nr}
at $\sqrt{s}=1.96\, \text{TeV}$.

Several attempts have been made to interpret these data, including QCD vacuum effects~\cite{bnm93,bbnu}
and higher-twist mechanisms~\cite{bbkd94,ehvv94}. In Ref.~\cite{boer} Boer demonstrated
that the product of two transverse momentum dependent (TMD) Boer-Mulders functions
$h_1^\perp(x,\boldsymbol{p}^2)$~\cite{bm} can produce unsuppressed $\cos 2 \phi$ asymmetries
that correspond to a violation of the Lam-Tung relation. Several theoretical and phenomenological
studies\cite{lm04,Lu:2005rq,
radici05,sissakian05,sissakian06,Lu:2006ew,Barone:2006ws,Lu:2007kj,
lms07,gamberg072,blms08a,Barone:2010gk}
along this direction have been put forward. Those studies are mainly concentrated on the $Q$ region much lower than the $Z$ mass,
where the lepton pair is produced via a virtual photon. In this paper, we will study the
phenomenology of the $\cos 2\phi$ asymmetry in the $Z$-pole region. We will show that the behavior of
the $\cos 2\phi$ asymmetry coming from the Boer-Mulders effect in the $Z$ mass region is very different
from that in the low $Q$ region.

\section{Description of $\cos2\phi$ asymmetries in terms of Boer-Mulders functions}

\begin{figure}
\begin{center}
\scalebox{1.2}{\includegraphics{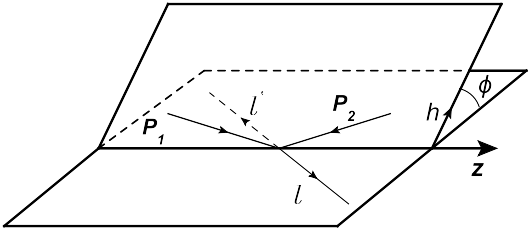}}\caption{\small  Angular
definitions of unpolarized Drell-Yan process in the lepton pair
center of mass frame.}\label{drell-yan}
\end{center}
\end{figure}

The process we consider here is the dilepton production via a $\gamma^*/Z$ boson in the unpolarized hadron collision:
\begin{eqnarray}
h_1 (P_1) + h_2 (P_2) \rightarrow  \gamma^*/Z (q) + X \rightarrow \ell^+ (\ell) + \ell^- (\ell^\prime) +X\,.\label{dipro}
\end{eqnarray}
The angular distribution coefficients $\lambda, \mu, \nu$ (or $W_{T,L,\Delta,\Delta \Delta}$) in this process
are generally frame dependent. In the following we will use
the Collins-Soper (CS) frame~\cite{cs77}, as shown in Fig.~\ref{drell-yan}. This frame is the c.m.s of the dilepton,
and in addition the z-axis is chosen to be along the bisector of
momenta $\boldsymbol{P}_1$ and $-\boldsymbol{P_2}$. In principle one can also choose the Gottfried-Jackson (GJ)
frame~\cite{lt78}. The advantage of CS frame in our study is that in this frame the Lam-Tung relation is rather insensitive
to higher-order corrections~\cite{mo94} and resummation effects~\cite{bv06}. Besides, it is more widely
used in theoretical and experimental studies, so that the comparison with other works is straight forward.

In the TMD factorization framework, the $\cos 2 \phi$ angular dependence arises from the coupling
of the Boer-Mulders functions, which depend on the intrinsic transverse momentum, and appear in the decomposition of the
TMD parton correlation function for an unpolarized hadron~\cite{bm}:
\begin{equation}
\Phi(x,\boldsymbol{p}_T)=\frac{1}{2}\left
[f_{1}(x,\boldsymbol{p}_T^2)n\ssh _++h_{1}^\perp(x,\boldsymbol{p}_T^2)\frac{\sigma_{\mu\nu}\boldsymbol{p}_T^\mu
n_+^{\nu}}{M}\right ].
\end{equation}
Here $n_+ = (0,1,\bm{0}_T)$ is a light-like vector expressed in the light-cone coordinates,
in which an arbitrary four-vector $a$ is written as $\{a^-,a^+,
\boldsymbol{a}_T\}$, with $a^{\pm}=(a^0 \pm a^3)/\sqrt{2}$ and $\boldsymbol{a}_T
=(a^1,a^2)$. The Boer-Mulders function describes the transverse polarization of
the quark inside a unpolarized nucleon, and thus is chiral-odd. Despite its time-reversal odd nature, $h_1^\perp$ can be nonzero, due to
inital/final state interactions~\cite{bhs02,collins02,bbh03,belitsky,bmp03} between the
struck quark and the spectator of the nucleon. These studies have motivated the model calculations, as well as the lattice analysis, of the Boer-Mulders functions for nucleon~~\cite{gg02,pobylista,yuan,bsy04,
Burkardt:2007xm,gamberg07,gpdtmd,Bacchetta:2008af,Wakamatsu:2009fn,Courtoy:2009pc,Bacchetta:2010si,Pasquini:2010af,Pasquini:2011tk,Gockeler:2006zu} and pion~\cite{lm04,Lu:2005rq,Meissner:2008ay,Gamberg:2009uk,Brommel:2007xd}

The detailed derivation of the angular dependent differential cross-section of reaction (\ref{dipro})
in the TMD factorization framework has been given in Refs.~\cite{boer,Arnold2009}. Here we write down
the final expression,
after taking into account both of photon and $Z$ boson contributions:
\begin{eqnarray}
&&\frac{d\sigma(h_1h_2\rightarrow l\bar{l}X)}{d\Omega
dx_ 1dx_2d^2\boldsymbol{q}_T}=\frac{\alpha^2}{3Q^2}\Bigg{\{}
K_1(\theta)F_{UU}^1\nonumber\\
&&\left.\,+\,[K_3(\theta)\cos2\phi+K_4(\theta)\sin2\phi] F_{UU}^{2\phi}\right\},\label{csbm}
\end{eqnarray}
in which only the unpolarized production is included.
In the above equation there are two structure functions contributing to the cross-section, which have the form:
\begin{eqnarray}
F_{UU}^1 &=&\sum_{q,\bar{q}}{\cal F}\left[f_1^{\,q}f_1^{\,\bar{q}}\right],\label{fuu1}\\
 F_{UU}^{2\phi}&=&\sum_{q,\bar{q}}{\cal F}\left [\left(2\hat{\boldsymbol{h}}\cdot
\boldsymbol{p}_T\, \hat{\boldsymbol{h}}\cdot \boldsymbol{k}_T
-\boldsymbol{p}_T\cdot
\kp\right)\frac{h_1^{\perp\,q}h_1^{\perp\,\bar{q}}}{M_1M_2}\right
].\label{bbpro}
\end{eqnarray}
 The vector
$\hat{\boldsymbol{h}}=\boldsymbol{q}_T/Q_T$.
, and we have used the
notation
\begin{eqnarray}
\mathcal{F}[\cdots]=\int d^2\boldsymbol{p}_T
d^2\kp\delta^2(\boldsymbol{p}_T+\kp-\boldsymbol{q}_T)
[\cdots].
\end{eqnarray}
Therefore (\ref{fuu1}) and (\ref{bbpro}) represent tree-level parton model results.

The coefficients $K_i$ in front of structure functions have the form~\cite{boer}
\ba
K_1(\theta)&=& {1\over 4}(1+\cos^2\theta)\left[ e_a^2+ 2 g_V^l e_a g_V^a \chi_1 + c_1^l c_1^a \chi_2
\right]
\nonumber\\
&+& \frac{\cos\theta}{2} \left[ 2 g_A^l e_a g_A^a \chi_1 + c_3^l c_3^a \chi_2
\right],\\
K_3(\theta)&=& {1\over 4}\sin^2\theta\left[ e_a^2+ 2 g_V^l e_a g_V^a \chi_1 + c_1^l c_2^a \chi_2
\right], \label{k3exp}\\
K_4(\theta)&=& {1\over 4}\sin^2\theta\left[ 2 g_V^l e_a g_A^a \chi_3 \right]
\ea
which contain the combinations of couplings
\begin{eqnarray}
c_1^j &=&\left(g_V^j{}^2 + g_A^j{}^2 \right),
\\[2 mm]
c_2^j &=&\left(g_V^j{}^2 - g_A^j{}^2 \right),
\qquad\qquad j=\ell\;\;\mbox{or}\;\;a
\\[2 mm]
c_3^j &=&2 g_V^j g_A^j.
\end{eqnarray}
The vector and axial-vector couplings to the $Z$ boson are given by:
\begin{eqnarray}
g_V^j &=& T_3^j - 2 \, Q^j\,\sin^2 \theta_W,\\
g_A^j &=& T_3^j,
\end{eqnarray}
where $Q^j$ denotes the charge and $T_3^j$ the weak isospin of
particle $j$ (for example, $T_3^j=+1/2$ for $j=u$ and $T_3^j=-1/2$ for
$j=e^-,d,s$).
The $Z$-boson propagator factors are given by
\ba
\chi_1 &=& \frac{1}{\sin^2 (2 \theta_W)} \, \frac{Q^2
(Q^2-M_Z^2)}{(Q^2-M_Z^2)^2 + \Gamma_Z^2 M_Z^2},\\
\chi_2 &=& \frac{1}{\sin^2 (2 \theta_W)} \, \frac{Q^2}{Q^2-M_Z^2} \chi_1,\\
\chi_3 &=& \frac{-\Gamma_Z M_Z}{Q^2-M_Z^2} \chi_1.
\ea

The first term in Eq.~(\ref{csbm}) is azimuthal independent. It gives the  $(1+\cos^2\theta)$ and $\cos\theta$ angular dependence while the later one vanishes after integration upon the polar angle $\theta$. The structure function $F_{UU}^1$ therefore corresponds to $W_T$ given in (\ref{acs}).
The second term has a $\cos2\phi$ azimuthal dependent term which
contributes to the asymmetry $\nu$~\footnote{In principle there can be also a $\sin2\phi$ azimuthal dependence.
However, it is $1/Q^2$ suppressed compared to the
 $\cos2\phi$ dependence and can be ignored here.}.
As shown in (\ref{bbpro}), it arises from the product of the transverse momentum dependent functions $h_1^\perp$ from each hadron.

One important feature implied by (\ref{bbpro}) is that the $\cos 2\phi$ dependence contributed by the
Boer-Mulders effect shows up in the pure electro-weak process without the need of QCD radiation. It will give a sizable contribution to $W_{\Delta\Delta}$ at low $q_T$.  Since the same effect cannot contribute to $W_L$, its presence violates the
Lam-Tung relation in the low $q_T$ region. If we express the structure function $W_{\Delta\Delta}$ as the sum of
the perturbative QCD effect and the Boer-Mulders effect
\begin{equation}
W_{\Delta\Delta} =W_{\Delta\Delta}^{QCD}+W_{\Delta\Delta}^{BM}
\end{equation}
where $W_{\Delta\Delta}^{BM}$ is proportional to $F_{UU}^{2\phi}$,
then the combination of the coefficients yields
\begin{eqnarray}
2\nu+\lambda-1 &=& {4(W_{\Delta\Delta}^{QCD}+W_{\Delta\Delta}^{BM}) \over W_T+W_L}
+ {W_T-W_L \over W_T+W_L}
-1 \nonumber\\
&\approx& {4W_{\Delta\Delta}^{BM} \over W_T+W_L} =2\nu^{BM}. \label{ltvio}
\end{eqnarray}
Here $\nu^{BM}$ denotes the $\cos 2\phi$ asymmetry contributed by the Boer-Mulders effect, in analogy with
the definition of $\nu$ in (\ref{coedif}).
In the above equation we have used the Lam-Tung relation
\begin{eqnarray}
2W_{\Delta\Delta}^{QCD} - W_L \approx 0
\end{eqnarray}
The $\approx$ sign follows the fact that higher order perturbative contributions still give a very small contribution. As explained previously, this contribution can be minimized by choosing the CS frame.
At low $q_T$, there is $W_L\ll W_T$, then we can arrive at following
expression for $\nu^{BM}$:
\begin{eqnarray}
\nu^{BM} (q_T,y,Q) = {2W_{\Delta\Delta}^{BM} \over W_T}\approx  {2F_{UU}^{2\phi}\over F_{UU}^1} \label{nubm}
\end{eqnarray}
The approximation in the above equation comes from the tree-level expressions (\ref{fuu1},\ref{bbpro}) for  $F_{UU}^1$ and $F_{UU}^{2\phi}$, that is, we do not consider the soft factor in the TMD factorization formula. We will comments on this approximation in the next section.

An important feature exposed by (\ref{k3exp}) is that the $\cos2\phi$ angular dependence contributed via the $Z$ boson has an opposite sign compared with the one contributed via a virtual photon. This can been seen from the fact that $e_a^2>0$,  while
\begin{equation}
c_2^a = \left(g_V^a{}^2 - g_A^a{}^2 \right)<0 ~~~~~~\textrm{for~~all~~flavors}.
\end{equation}
The minus sign in the above equation comes from the odd permutation of gamma matrices in
 calculating the hadronic tensor coming from the Boer-Mulders effect:
\begin{equation}
\text{Tr}\left( \sigma_{\alpha\beta}
\,V^\mu \, \sigma_{\rho\sigma}
 \, V^\nu \right)
\end{equation}
where $V^\mu = g_V \gamma^\mu + g_A \gamma_5 \gamma^\mu$
denotes the $Z$-boson-fermion vertex.

At low $Q$ the dilepton production via virtual photon dominates. However, as
the dilepton mass approaches the $Z$ boson mass, the production via $Z$-boson becomes important and dominates over
that via virtual photon (the second term in (\ref{k3exp}) is suppressed in the $Z$ mass region compared with the third term). Thus one expects to observe the sign reversal of $\nu^{BM}$ (or, equivalently, the
combination $2\nu+\lambda-1$ that can be measured by experiment) from the low $Q$ region
to the $Z$-pole region. The sign reversal of $\nu^{BM}$ is due to the chiral-odd nature of the Boer-Mulders function,
and is a special feature of the Boer-Mulders effect as a source of the Lam-Tung relation violation.
Therefore, the experimental detection of the sign reversal of $2\nu+\lambda-1$ is a clear evidence on the existence of chiral-odd quarks inside the unpolarized nucleon. The experiments can be conducted at hadron colliders with unpolarized beams, and thus this allows to study the spin structure of the nucleon without polarized beams.

\begin{figure*}
\begin{center}
\scalebox{0.7}{\includegraphics*[25pt,15pt][315pt,260pt]{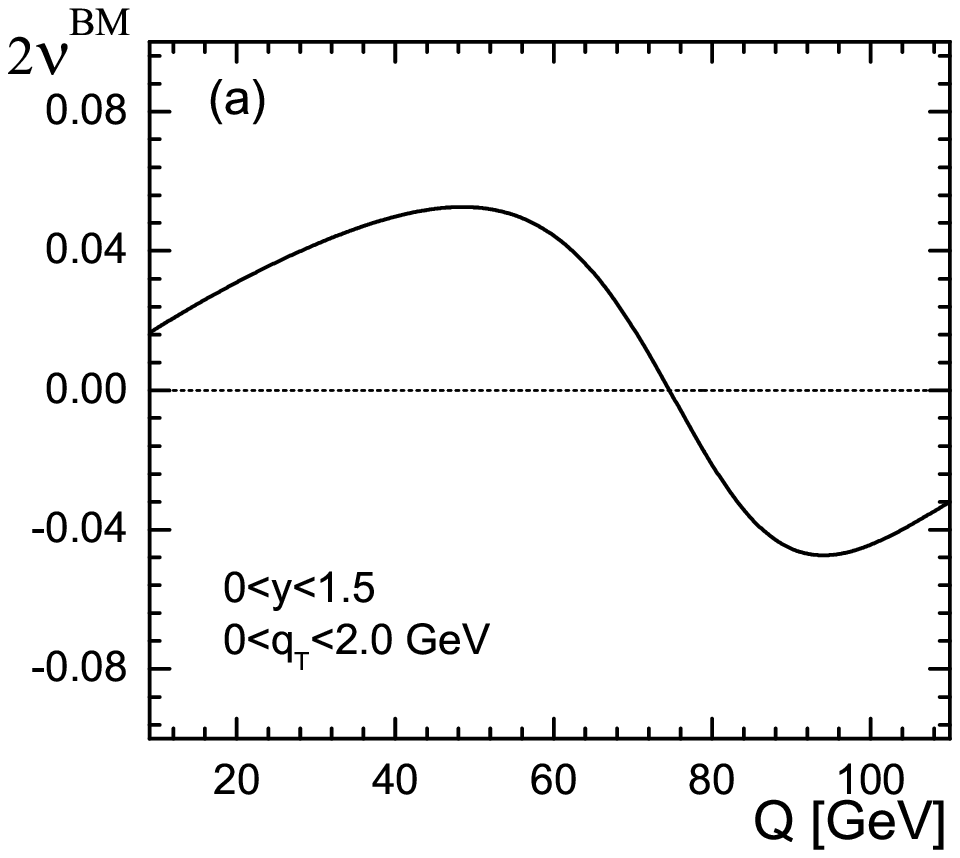}}
\scalebox{0.7}{\includegraphics*[25pt,15pt][315pt,260pt]{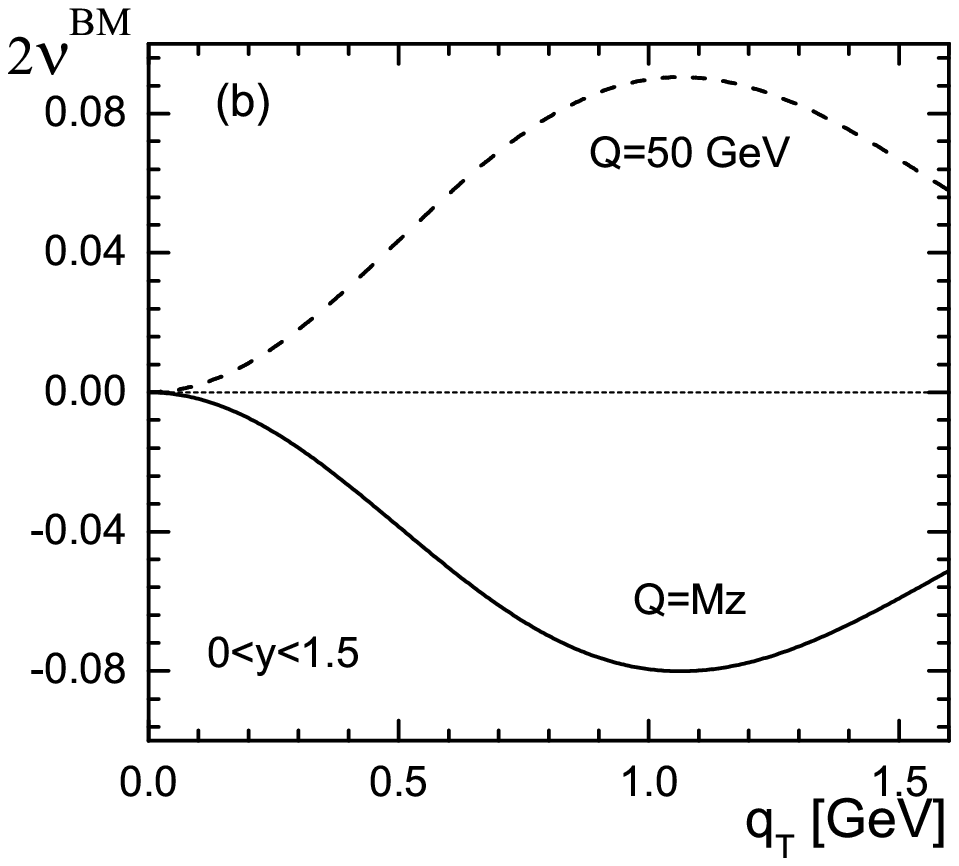}}
\caption{\small (a): The $Q$-dependent $\cos 2\phi$ asymmetry in $pp\rightarrow \gamma^*/Z+ X \rightarrow \ell^+\ell^+-X$ process at RHIC. (b): The $q_T$-dependent $\cos 2\phi$ asymmetry in the unpolarized $pp\rightarrow \gamma*/Z+ X \rightarrow \gamma^*/Z+ X$ process at RHIC for different $Q$ values  }\label{rhic-q-qt-dep}
\end{center}
\end{figure*}

\begin{figure*}
\begin{center}
\scalebox{0.7}{\includegraphics*[25pt,15pt][315pt,260pt]{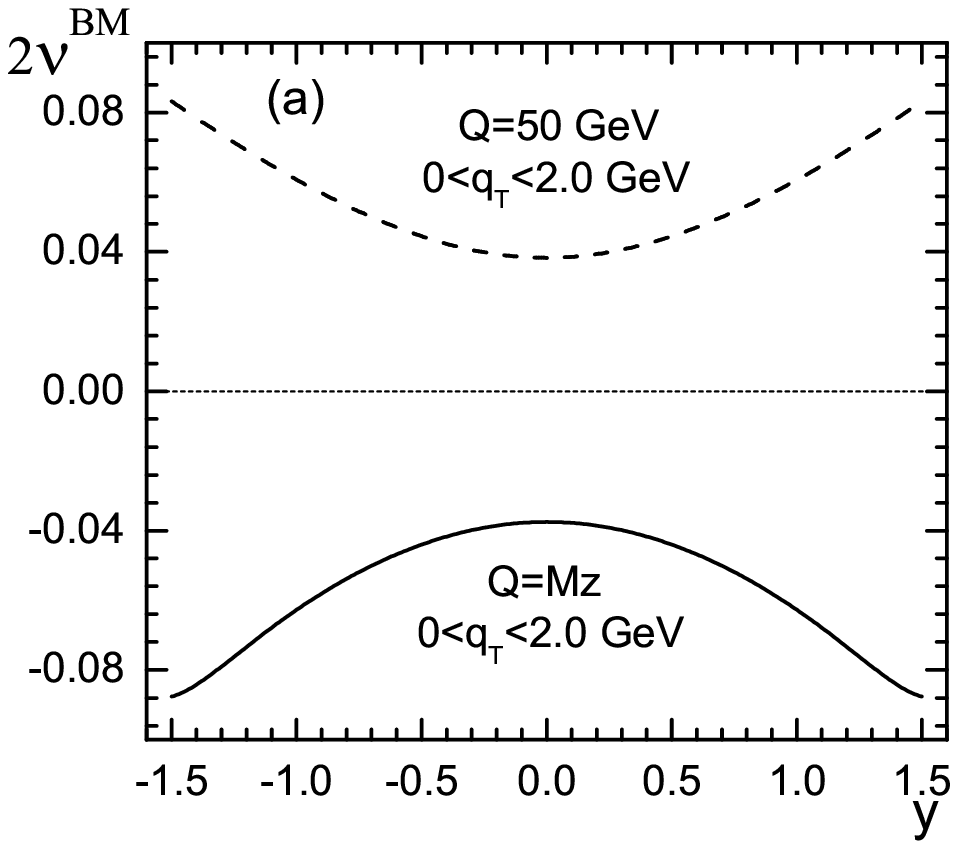}}
\scalebox{0.7}{\includegraphics*[25pt,15pt][315pt,260pt]{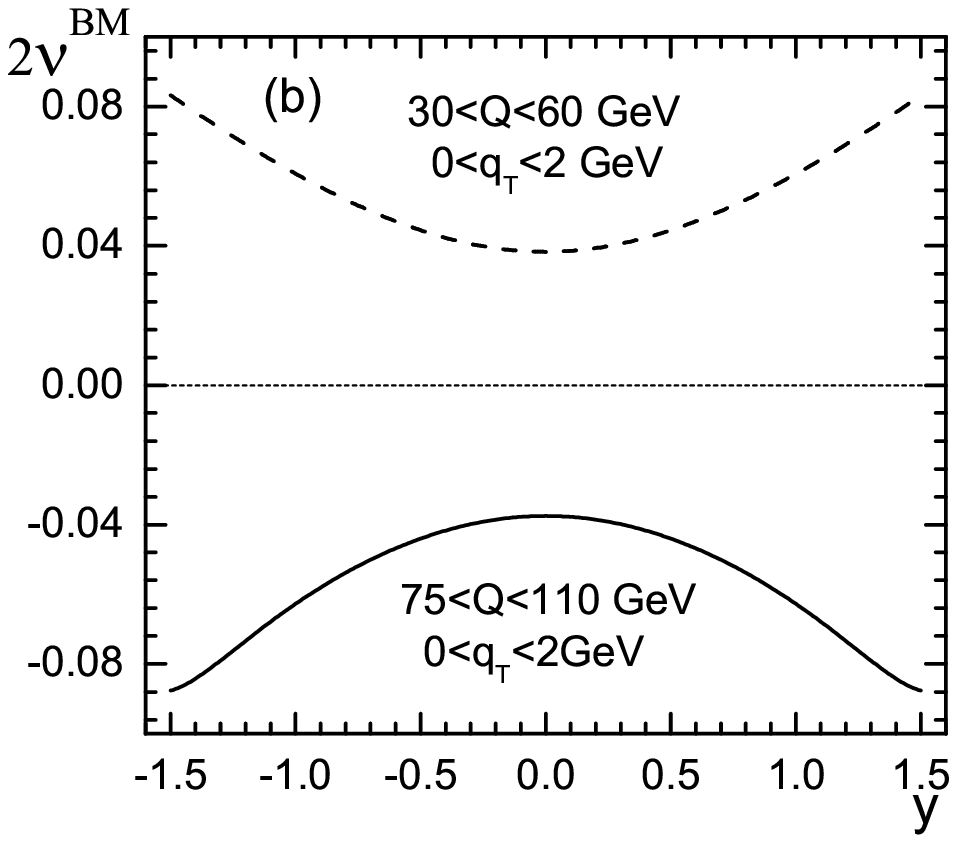}}
\caption{\small (a): The $Q$-dependent $\cos 2\phi$ asymmetry in $pp\rightarrow \gamma^\star/Z+ X \rightarrow \ell^+\ell^-+X$ process at RHIC. (b): The $q_T$-dependent $\cos 2\phi$ asymmetry in the unpolarized $pp\rightarrow \gamma^*/Z +X \rightarrow \ell^+\ell^-X$ process at RHIC for different $Q$ values  }\label{rhic-y-dep}
\end{center}
\end{figure*}

\section{Numerical Results}

We will show that RHIC~\cite{Bunce:2000uv} at BNL is ideal for measuring the sign-reversal of $\nu^{BM}$.
The $pp$ Drell-Yan process in the RHIC-spin program is mainly dedicated to polarized proton beams, although a spin-averaged measurement is still allowed.
At RHIC, with the highest center of mass energy $\sqrt{s}=500 \text{GeV}$,
the experiments can probe the region $x \sim 0.2$ (corresponding to the mid-rapidty of the vector boson)
as the dilepton mass is near the $Z$-pole.
We will present the prediction for $2\nu^{BM}$, since it is equal to $2\nu+\lambda-1$, and the later is the observable that can be directly measured by experiments.

For the Boer-Mulders functions needed in the calculation, we adopt the parametrization~\cite{Lu:2009ip} extracted from the unpolarized $pd$ and $pp$ Drell-Yan data at  $0<q_T<2~\textrm{GeV}$ measured by E866/NuSea Collaboration at FNAL. The E866/NuSea experiment covers the region $4.5<Q < 15~\textrm{GeV}$ (excluding the $\Upsilon$ resonance
region) with a center of mass energy $\sqrt{s}=38.7~\text{GeV}$. Therefore the $x$ region near the $Z$-pole at RHIC is similar to that of  the E866/NuSea experiment. Also RHIC is a proton-proton collider where the dilepton is produced by the annihilation of valence and sea quark from each hadron, just like the case of the E866/NuSea experiments.
The kinematical cuts applied in the calculations are:
\begin{eqnarray}
-1.5<y<1.5,~~~0<q_T<2 \textrm{GeV},
\end{eqnarray}
where $y={1\over 2}\ln(x_1 / x_2)$ is the rapidity of the $\gamma^*$ or $Z$ boson. The reasons to choose a low $p_T$ cut are twofold.  First at low $q_T$ the corrections from QCD are small, therefore the approximation in (\ref{nubm}) is valid. The other is that in at this special low $q_T$ region the intrinsic
transverse momentum of partons is more relevant, thus the tree-level result for $\nu^{BM}$ is justified.

In Fig.~\ref{rhic-q-qt-dep}a we plot the prediction for $\nu^{BM}$ (scaled with a factor of 2, in order to correspond to the size of the Lam-Tung relation violation) as a function of dilepton mass $Q$ at RHIC for $\sqrt{s}=500~\text{GeV}$. The result clearly shows that the value of $2\nu^{BM}$ is positive as $Q < 60~\text{GeV}$,  while it reverses sign to be negative as $Q>75~ \textrm{GeV}$, and reaches $-5\%$ at the $Z$-pole. In Fig~\ref{rhic-q-qt-dep}b we plot the prediction for $2\nu^{BM}$ as a function of transverse momentum of the dilepton  $q_T$, for $Q=50~\text{GeV}$ and $Q=M_Z$. Again it shows a sign reversal of $2\nu^{BM}$ at two different $Q$ value. The magnitude of the $2\nu^{BM}$ peak at $Q_T=1 \text{GeV}$  falls  at higher $q_T$. The $q_T$ shape of the asymmetries  indicate that the intrinsic transverse momentum of the parton plays a significant role at low $q_T$, and can give a substantial contribution. In Fig.~\ref{rhic-y-dep}a we plot the $y$-dependent results of $2\nu^{BM}$ for $Q=50~\text{GeV}$ and $Q=M_Z$, and in Fig.~\ref{rhic-y-dep}b we plot the $y$-dependent results of $2\nu^{BM}$ for $30<Q<45~\text{GeV}$ and $75<Q<110~\text{GeV}$. An observation from these two figures is that the magnitude of $2\nu^{BM}$ increases as the rapidity increases.
Our theoretical predictions suggests that the magnitude of $2\nu+\lambda-1$ is sizable near the $Z$-pole and is measurable at RHIC. Therefore an accurate measurement on $2\nu+\lambda-1$ both at $Z$-pole and at a lower dilepton mass region can serve as a test of the chiral-odd property of quarks inside an unpolarized nucleon.

Several points need to be addressed here. First, as the evolution~\cite{Henneman:2001ev} of the TMD Boer-Mulders function still remains unclear, we assume that the scale dependences of $h_1^\perp(x,\boldsymbol{p}_T^2)$ and
the spin averaged distribution function $f_1(x,\boldsymbol{p}_T^2)$ are the same in calculating $\nu^{BM}$,
which was also adopted before in the extraction of the Boer-Mulders function~\cite{Lu:2009ip}.
We would like to admit that the evolution of $h_1^\perp(x,\boldsymbol{p}_T^2)$ can be more complicated than
that of $f_1(x,\boldsymbol{p}_T^2)$, as the former one is chiral-odd, while the later one is chiral-even.
Further more, recent quantitative calculation~\cite{Aybat:2011zv} on the spin-independent processes demonstrate that the Collins-Soper evolution of TMD distributions may be significant.
 Since $\nu^{BM}$ is
approximately the ratio between $h_1^\perp$ and $f_1$, the evolution effect can
only influence our results quantitatively at most, but not qualitatively. As our main purpose is to
reveal the sign change of $\nu^{BM}$ between the low $Q$ region and $Z$ mass region, our assumption
on the scale dependence can be viewed as a reasonable choice. Secondly, in Eqs.~(\ref{fuu1},\ref{bbpro})
we employ a tree-level expression of the TMD factorization formula, in which the soft factor has not been considered.
Resummation effects~\cite{Collins:1992kk,Boer:2001he,Kang:2011mr} of soft gluon radiation in the
TMD factorization will lead to a Suddakov factor that alters the tree-level result. The study in
Ref.~\cite{Boer:2001he} shows that TMD azimuthal spin asymmetries are suppressed by this Suddakov
factor in the region where $q_T$ is much larger than the intrinsic transverse momentum of the parton,
but still much smaller than $Q$. In our calculation we restrict the cut on the transverse momentum of
the dilepton as $0<q_T<2~\text{GeV}$, where the intrinsic transverse momentum of partons plays a significant role,
and we assume that the tree level approximation still holds, to avoid Suddakov suppression. The Suddakov
effect is certainly important for azimuthal observables at higher $q_T$ (but still much small than $Q$)
and should be considered. Based on the uncertainties discussed above, our result can be viewed as an estimate.
Nevertheless, our study provides a useful understanding of the $\cos 2\phi$ azimuthal asymmetry in the $Z$ mass region
from the tree-level calculation. Very recently, new theoretical analysis given in Ref.~\cite{Boer:2011xd} shows that, soft factors appearing beyond tree level cancel out of the weighted azimuthal asymmetry by employing Bessel functions. We expect that this newer approach can be applied to study the weighted $\cos 2\phi$ azimuthal asymmetry at the $Z$-pole to provide a rigorous test on the sign change of $\nu^{BM}$.

\section{Conclusion}

We study the $\cos 2\phi$ angular dependence of dilepton production at the $Z$-pole. We show that, due to the chiral-odd nature of Boer-Mulders function, the Boer-Mulders effect will cause a sign change of $2\nu +\lambda-1$ (or equivalently, $2\nu^{BM}$) from the low $Q$ region to the $Z$ mass region. This is a special signature of the Boer-Mulders effect as a source of violation of the Lam-Tung relation. Therefore, the experimental detection of the sign reversal of $2\nu+\lambda-1$ will serve as a clear evidence of the existence of chiral-odd quarks inside an unpolarized nucleon.
Using a recent extracted set of Boer-Mulders functions, we predict the sign and magnitude of $\nu^{BM}$ near the $Z$-pole at RHIC with $\sqrt{s}=500\,\text{GeV}$, based on TMD factorization. Our analysis at the low $Q$ and $Z$ mass regions shows that this sign reversal can be detected if the Boer-Mulders effect indeed is a source of Lam-Tung relation violation. A dedicated measurement on both of the size and the sign of the Lam-Tang violation at RHIC in a wide $Q$ range therefore provides valuable hint on the dynamics of the $\cos2\phi$ azimuthal dependence in dilepton production.

{\bf Acknowledgements} This work is supported by FONDECYT (Chile) Project Nos.~11090085,
No.~1100715, by Project Basal FB0821,
 and by NSFC (China) Project No.~11005018.

\end{document}